\documentclass[sn-mathphys,Numbered]{sn-jnl}

\usepackage{graphicx}%
\usepackage{multirow}%
\usepackage{amsmath,amssymb,amsfonts}%
\usepackage{amsthm}%
\usepackage{mathrsfs}%
\usepackage[title]{appendix}%
\usepackage{xcolor}%
\usepackage{textcomp}%
\usepackage{manyfoot}%
\usepackage{booktabs}%
\usepackage{algorithm}%
\usepackage{algorithmicx}%
\usepackage{algpseudocode}%
\usepackage{listings}%

\begin{document}

\title{Reassessing the impact of megaelectronvolt ions in fusion plasmas via gyrokinetic simulations}

\author*[1]{\fnm{Alessandro} \sur{Di Siena}}\email{alessandro.di.siena@ipp.mpg.de}
\equalcont{These authors contributed equally to this work.}

\author*[2]{\fnm{Gabriele} \sur{Merlo}}\email{gmerlo@ices.utexas.edu}
\equalcont{These authors contributed equally to this work.}

\author[1]{\fnm{Alejandro} \sur{Ba\~n\'on Navarro}}\email{alejandro.banon.navarro@ipp.mpg.de}

\author[1]{\fnm{Tobias} \sur{G\"orler}}\email{tobias.goerler@ipp.mpg.de}

\author[1,3]{\fnm{Frank} \sur{Jenko}}\email{frank.jenko@ipp.mpg.de} 

\affil[1]{\orgname{Max Planck Institute for Plasma Physics}, \orgaddress{\city{Garching bei Munchen}, \country{Germany}}}

\affil[2]{\orgdiv{Oden Institute for Computational Engineering and Sciences}, \orgname{The University of Texas at Austin}, \orgaddress{\city{Austin}, \postcode{78712}, \state{Texas}, \country{USA}}}

\affil[3]{\orgdiv{School of Computation, Information and Technology}, \orgname{Technical University of Munich}, \orgaddress{\city{Munich}, \country{Germany}}}

\maketitle

Gyrokinetic simulations conducted by Mazzi et al.~\cite{Mazzi_Nature_2022} reveal the suppression of turbulence in fusion plasmas through the destabilization of Toroidal Alfv\'en Eigenmodes (TAEs) by megaelectronvolt ions. In their investigation of recent JET experiments, the authors explore the phenomenon of high ion temperatures and reduced density fluctuations in the presence of fast-ion-driven TAEs, as detected via the analysis of magnetic coil spectrograms and of Fourier spectra obtained from reflectometer measurements. When performing flux-tube gyrokinetic simulations, the authors uncover a noteworthy decrease in the heat conductivity of the thermal species when fast-ion-driven TAEs are strongly unstable, thereby reproducing the power balance derived from TRANSP calculations. This finding presents an intriguing departure from the typical behavior observed in electron turbulent fluxes, where the presence of unstable fast ion modes generally leads to confinement degradation \cite{Gorelenkov_NF_2010,Citrin_PPCF_2014,Doerk_PPCF_2016,DiSiena_NF_2019,DiSiena_JPP_2021}.

With this manuscript, we are not questioning the experimental results presented in the paper, but the authors interpretation of the observations on the basis of the gyrokinetic GENE simulations. In particular, our analysis demonstrates that the authors' numerical findings are strongly influenced by the selected simulation settings, calling into question their claim that the resulting heat conductivity aligns with the TRANSP power balance. Specifically, we assert that the numerical results presented by Mazzi et al.~\cite{Mazzi_Nature_2022} are a direct consequence of the inadequacy of resolution employed in their numerical simulations. Notably, there are three primary factors contributing to this issue: (i) the employed radial box size is insufficient, leading to an undesirable impact of the boundary conditions on the simulations; (ii) the adoption of a higher minimum toroidal mode number fails to accurately resolve the entire range of TAEs, resulting in an underestimation of TAE drive and introducing spurious effects on wave-particle resonances; (iii) an insufficient resolution in the magnetic moment direction exacerbates these challenges.

The lack of numerical convergence in the simulations conducted by Mazzi et al.~\cite{Mazzi_Nature_2022} is demonstrated in this study using the same gyrokinetic code (GENE \cite{Jenko_PoP2000}) and identical input parameters. Initially, we replicated their simulations using the same parameters to those provided by the authors for a specific case with $R/L_{p_{FD}} \approx 14$. Subsequently, we improved the numerical resolution in various directions. A comparison of the resolution employed in Ref.~\cite{Mazzi_Nature_2022} and the increased resolution setup utilized in our simulations is summarized in Table \ref{tab1}.
\begin{table}[h]
\caption{The table presented here outlines the variations in numerical setups between the nonlinear simulations described in Ref.~\cite{Mazzi_Nature_2022} and the simulations performed with higher resolution. Please refer to the "Methods" section of Ref.~\cite{Mazzi_Nature_2022} for the definitions of the quantities listed. All numerical parameters not specified in the table are assumed to be the same as those used in Ref.~\cite{Mazzi_Nature_2022}.}\label{tab1}
\begin{tabular}{@{}lll@{}}
\toprule
numerical parameters & Resolution Ref.~\cite{Mazzi_Nature_2022}  & High resolution setup \\
\midrule
$n_{k_x}$    & 256   & 512  \\
$n_{k_y}$    & 48   & 128  \\
$n_{0,min}$    & 4  & 1  \\
$n_{w0}$    & 64   & 30  \\
$\mu$ discretization    & Gauss-Legendre   & Equidistant \footnotemark[1]  \\
$l_x$    & 264   & 556  \\
$m_{ref}$    & 1   & 2  \\
\botrule
\end{tabular}
\footnotetext[1]{The use of an equidistant grid in $\mu$ allows to reduce the resolution and ensure numerical convergence. This is due to the high-$\mu$ contributions given by the wave-particle resonant interactions.}
\end{table}
Upon addressing the aforementioned numerical issues, a significant increase in heat conductivity for each plasma species was observed, as detailed in Table \ref{tab2}, diverging from the expected values derived from the TRANSP power balance calculations.
\begin{table}[h]
\caption{Comparison of the heat conductivity obtained from the low resolution setup employed in Ref.~\cite{Mazzi_Nature_2022}, the high resolution setup of Table \ref{tab1} and the TRANSP power balance for each thermal plasma species.}\label{tab2}%
\begin{tabular}{@{}llll@{}}
\toprule
Plasma species & Resolution Ref.~\cite{Mazzi_Nature_2022}  & High resolution setup & TRANSP \cite{Mazzi_Nature_2022}\\
\midrule
Deuterium (D) &  $\chi_D = 0.3m^2/s$   & $\chi_D = 1.74m^2/s$  & $\chi_D = 0.8m^2/s$  \\
Helium ($^3$He) &   $\chi_{^3He} = 0.3m^2/s$   & $\chi_{^3He} = 3.4 m^2/s$  & $\chi_{^3He} = 0.8m^2/s$  \\
Electrons (E)   & $\chi_E = 0.7m^2/s$   & $\chi_E = 7.2 m^2/s$  & $\chi_E = 0.5m^2/s$  \\
\botrule
\end{tabular}
\end{table}
Remarkably, the electron conductivity surpasses the values obtained from TRANSP by more than tenfold, consistent with earlier findings extensively documented in the existing literature \cite{Citrin_PPCF_2014,Doerk_PPCF_2016,DiSiena_NF_2019,DiSiena_JPP_2021}. These results definitively demonstrate that simulations incorporating strongly unstable TAEs are unable to reproduce the results achieved through TRANSP, thus decisively refuting the authors' claim of enhanced performance in the presence of such TAEs.

To present a comprehensive overview of our findings, we address the the critical issues of the simulations presented in Ref.~\cite{Mazzi_Nature_2022} below:

\textit{i) Insufficient radial box size} - We begin by highlighting the influence of the inadequate radial box size on the numerical results, emphasizing the successful mitigation of this issue through our high-resolution setup. Figure 1a illustrates the time evolution of the electron energy flux, computed by combining the electrostatic and electromagnetic components using the resolution specified in Ref.\cite{Mazzi_Nature_2022}. The computed value of $\chi_e = 0.7 m^2/s$ aligns with the reported findings in Ref.\cite{Mazzi_Nature_2022}. The figure also showcases contour plots of the perturbed electrostatic potential at different time snapshots. These plots provide clear evidence of turbulent eddies extending throughout the entire simulation box at each time slice, implying that the radial box size is inadequately small. Therefore, the boundary conditions \cite{Beer_PoP_1995} exert an unphysical influence on the observed results. Notably, when averaging over the saturated nonlinear phase, the perturbed electrostatic potential reproduces the structures depicted in Figure 4b of Ref.~\cite{Mazzi_Nature_2022}.
\begin{figure*}[h]%
\centering
\includegraphics[width=0.5\textwidth]{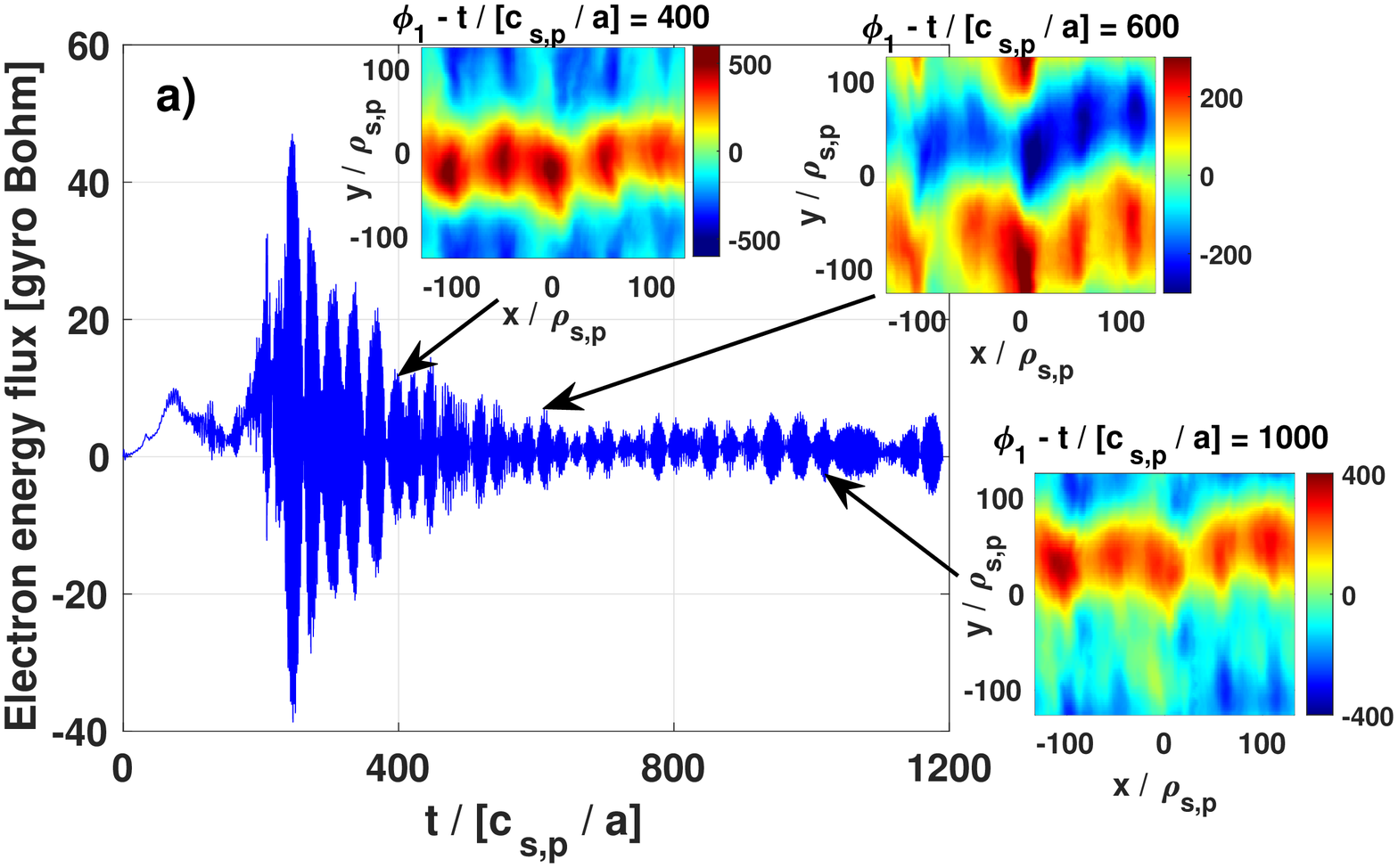}\includegraphics[width=0.5\textwidth]{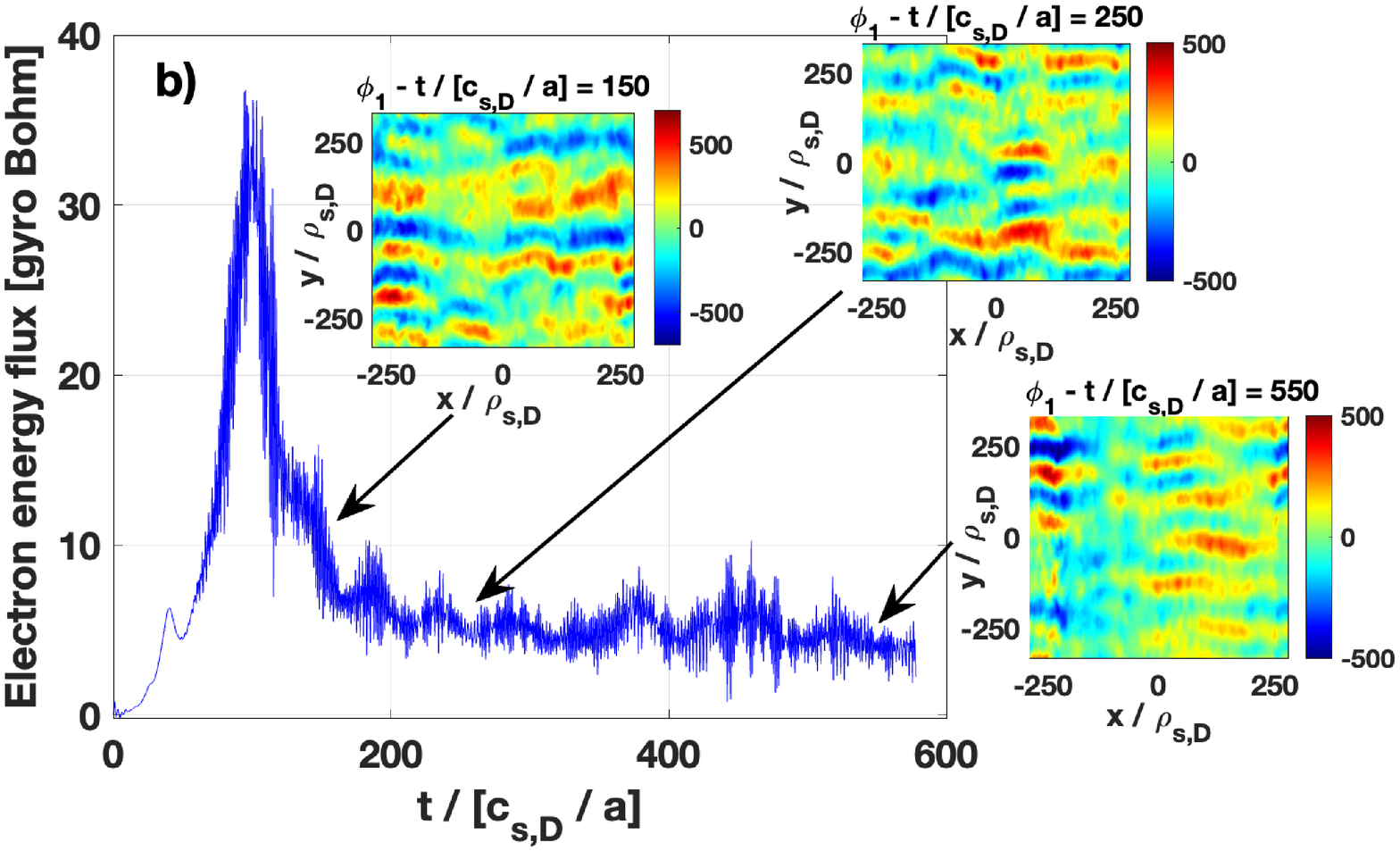}
\caption{Time evolution of the electron energy flux in gyro Bohm units for the a) low resolution setup employed in ref.~\cite{Mazzi_Nature_2022} and b) high resolution setup of Table \ref{tab1}. The insets represent the perturbed electrostatic potential $\phi_1$ at different time-slices of the simulations in the bi-normal ($y$) and radial ($x$) plane. The time is expressed in units of proton ($c_{s,p}$) and deuterium ($c_{s,D}$) sound speeds, respectively, for a) and b) and minor radius $a$, while $x$ and $y$ are normalized to the proton ($\rho_{s,p}$) and deuterium ($\rho_{s,D}$) sound Larmor radii, respectively, for a) and b).}
\label{fig1}
\end{figure*}
Figure 1b highlights the significant changes in the time evolution of the electron energy flux (and likewise for each plasma species) when as enhanced numerical resolution and expanded box sizes are used. The saturated heat conductivity reaches a value of $\chi_e = 7.2 m^2/s$, surpassing the TRANSP power balance by more than tenfold. The previously observed unphysical oscillations of large amplitude around zero are no longer observable. Furthermore, the contour plots of the perturbed electrostatic potential no longer portray turbulent eddies extending throughout the entirety of the box.

\textit{ii) Inadequate minimum toroidal mode number} - The resolution used in Ref.~\cite{Mazzi_Nature_2022} yields abnormal behavior in the energy flux spectra, as demonstrated in Figure 2 for the fast ion species. With this low resolution setup, the electrostatic spectra of fast ions exhibit a peak at the minimum finite toroidal mode number $(n_{0,min} = 4)$, indicating that the selected box size in the $y$ direction is also insufficient, and $n_{0,min}$ fails to accurately capture the turbulent structures at low toroidal mode numbers \footnotemark[2]. \footnotetext[2]{The toroidal mode number discretization in the GENE code is expressed as $n = n_{0,min} \cdot j$, where $j$ represents an integer value ranging from $j = [0,1,2, ..., n_{ky}]$}
A similar issue is observed in the spectra of the other thermal species, as depicted in Figure S7a of Ref.\cite{Mazzi_Nature_2022}. However, by reducing the toroidal mode number to $n_{0,min} = 1$, the linear spectra of TAEs are accurately resolved, as illustrated in Figure 2. This higher resolution setup effectively resolves the problematic feature in the energy flux spectra of fast ions and thermal species. The magnetic coil spectrograms of JET pulse $\#94701$ presented in Figure 2e of Ref.\cite{Mazzi_Nature_2022} also supports the use of $n_{0,min} = 1$, revealing the presence of unstable TAEs in the range of $n = [2 - 5]$. 
\begin{figure}[h]
\centering
\includegraphics[width=0.7\textwidth]{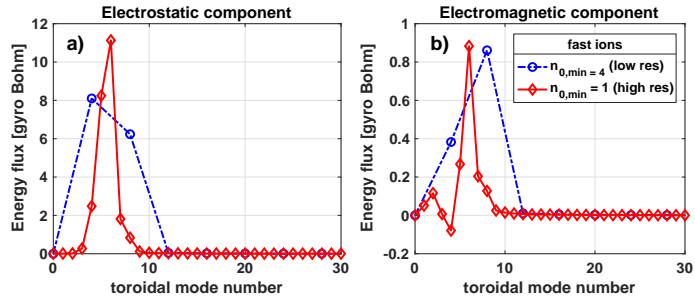}
\caption{Fast ion energy flux spectra in gyro Bohm units for the a) electrostatic and b) electromagnetic components obtained with the low resolution setup employed in Ref.~\cite{Mazzi_Nature_2022} (dashed) and the high resolution setup of Table \ref{tab1} (solid).}
\label{fig2}
\end{figure}

\textit{iii) Insufficient Magnetic Moment Resolution} - Another aspect of our argument revolves around the inadequately resolved velocity space structure of the fast ion energy flux resulting from the employed quadrature rule as outlined in Ref.~\cite{Mazzi_Nature_2022}. Figure 3a effectively highlights this concern, where the largest contribution of the fast ion to the energy flux is localized at the final grid point in the magnetic moment, artificially impacting the numerical results. Specifically, the use of Gauss-Legendre discretization, which inherently establishes a non-equidistant grid with a decreasing resolution at higher magnetic moments, leads to inadequate resolution of fast ion resonances at larger magnetic moments. However, when using an equidistant grid and our numerical setup, this problematic characteristic is successfully mitigated, as demonstrated in Figure 3b. Additionally, we observe a displacement of fast ion resonances within the velocity space when employing the high-resolution configuration. This result is likely associated with the under-resolved energy flux spectra presented in Figure 2a, which exclusively captures the $n = [4, 8]$ modes, consequently influencing the resonant conditions $f_{TAE} - n \Omega_{prec} = 0$ (where $\Omega_{prec}$ represents the toroidal precession frequency of trapped fast ions) necessary for the excitation of these TAEs \cite{Chen_RMP_2016}.

\begin{figure}[h]%
\centering
\includegraphics[width=0.7\textwidth]{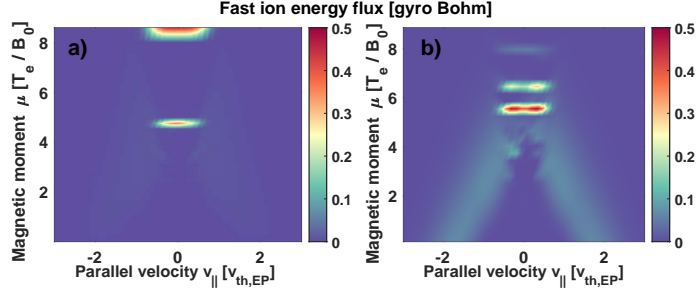}
\caption{Velocity space structure of the fast ion energy flux in gyro Bohm units for the a) low resolution setup employed in Ref.~\cite{Mazzi_Nature_2022} and b) the high resolution setup of Table \ref{tab1}.}
\label{fig3}
\end{figure}

In conclusion, the claims made by Mazzi et al.~\cite{Mazzi_Nature_2022} regarding the enhanced performance of fusion devices in the presence of megaelectronvolt ions and strong fast ion-driven Alfv\'enic activity are unfounded, as their simulations lack numerical convergence. Our investigations employing the high resolution setup confirm previous findings, demonstrating a significant increase in turbulent fluxes when strongly (linearly) unstable Alfv\'enic modes are present in flux-tube simulations. Consequently, our results raise serious doubts on the authors' assertion of enhanced performance in the presence of strongly unstable TAEs, emphasizing the need for a reevaluation of their claims.

\end{document}